# Photoinduced Topological Phase Transitions in Kitaev kagome ferromagnet

Hongchao Shi [1), a]   Zhengguo Tang [1), a]   Bing Tang [1)†]   Chaofei Liu[2)‡]

1)(School of Physics and Mechanical & Electrical Engineering, Jishou University, Jishou 416000, China)

2) (School of Science, Jiangxi University of Science and Technology, Ganzhou 341000, China)

The photoinduced topological phase transitions and thermal Hall conductivity of a kagome magnet with Heisenberg, Kitaev, and DM interactions under an external magnetic field aligned along the [111] directions is investigated in this study. In the presence of a strong magnetic field perpendicular to the lattice plane in the [111] direction, the system exhibits a fully polarized paramagnetic phase and the magnon band carries an asymmetric Chern number across the phase diagram region. Utilizing magnetic Floquet-Bloch theory, we demonstrate that periodically driven intrinsic topological magnetic materials can be manipulated into different topological phases with varying Berry curvature, Chern numbers and thermal Hall conductivities signatures by adjusting light intensity throughout the phase diagram region.

**1 Introduction**

In the last few decades, there has been a growing interest in studying the

† Corresponding author. E-mail: bingtangphy@jsu.edu.cn
‡ Corresponding author. E-mail: liuchaofei@jxust.edu.cn
a These authors contributed equally to this work topological properties of condensed matter systems, both theoretically and experimentally, and a prominent result is that band structures in solids can carry non-trivial topological indices, which determine and protect certain properties of the solid's spectrum at interfaces, exhibiting the quantum anomalous Hall effect due to topologically protected chiral edge states [1-5]. It is possible to extend this concept from electronic systems to bosonic systems, such as the magnon [6-15] Similar to electronic systems, topological magnonic materials also possess exotic topological properties, such as the non-zero Berry curvature responsible for the thermal Hall effect describing the charge-neutral spin excitations carrying heat flowing transversely under external magnetic fields, and edge or surface state that is topologically protected and robust to perturbations.

Indeed, in ordered magnets, topological magnon insulator has been investigated as a bosonic analog of the topological insulator, in the honeycomb lattice with the Dzyaloshinskii Moriya interaction (DMI) [16, 17] and many other systems, such as magnets with the Kitaev interaction [18, 19] and kagome magnet[20]. In real materials, the fact that Kitaev interactions are usually accompanied by DMI and that the same magnetic oscillator dispersion can be explained based on DMI, Kitaev interactions, or a combination of them, leads to the fact that it is often difficult to distinguish which one dominates the system. However, in the presence of an external planar magnetic field, if the dispersion of the system changes drastically due to a change in the direction of the magnetic field, the system is dominated by Kiaev interactions rather than DMI [21]. In addition, another indication of the dominance of the Kitaev interaction in the system can be provided by the sign changes in the thermal transverse properties as a function of temperature or the strength of an external magnetic field.

Essentially, the intrinsic properties of a particular topological magnet insulator are material constants that cannot be tuned, thus preventing topological phase transitions in the material. However, in many cases of physical significance, manipulating the intrinsic properties of topological magnetic materials can be a stepping stone to promising practical applications, and can also provide a platform for the study of new and interesting properties, e.g., photo-magnonics [22], magnon spintronics [23-24]. One of the ways to achieve these scientific goals is undoubtedly through light-matter interactions induced by light irradiation. In recent years, formalism of photo-irradiation has been the main focus of electronic systems (e.g., graphene, etc. e.g., graphene, etc.)[25-29], However, theoretically, in a similar manner to the notion of bosonic topological band theory, one can also extend the mechanism of photo-irradiation to bosonic systems.

In this work, we study the photo-irradiated intrinsic topological magnon insulators and associated topological phase transitions of kagome magnet exhibiting Heisenberg, Kitaev and DM interactions exposed to a magnetic field. We will use the quantum theory of magnon to achieve this. Magnon usually carry a magnetic dipole moment. Thus, magnons can couple via the Aharonov-Casher (AC) effect [26] to both time-independent and periodic time-dependent electric fields [27-31], much as electronic charged particles couple via the Aharonov-Bohm (AB) effect [32].

Interestingly, given periodic time-dependent electric fields, this results in a periodically driven magnon system, which may be analyzed using the Floquet-Bloch theory. Using this theory, we show that by manipulating the light intensity, intrinsic topological magnon insulators can be tuned from one topological magnon insulators to another with different Berry curvature, Chern number, and thermal Hall conductivity. As a result, it is therefore possible to regulate the magnon spin current in topological magnetic materials using photo-irradiation, which is expected to be validated in future practical applications.

## 2. Model and method

We consider the Kitaev-Heisenberg model on the kagome plane lattice, sketched in Fig. 1. The corresponding spin Hamiltonian is given by

$$\mathcal{H} = J\sum_{\langle i,j \rangle} \vec{S}_i \cdot \vec{S}_j + K \sum_{\langle i,j \rangle \gamma} S_i^\gamma S_j^\gamma + \sum_{\langle i,j \rangle} \vec{D}_{ij} \cdot \left( \vec{S}_i \times \vec{S}_j \right) - \vec{B} \cdot \sum_i \vec{S}_i \tag{1}$$

where, the first term is the nearest-neighbor Heisenberg exchange interaction, the second term is the bond-dependent Kitaev interaction, these can parameterized as $J = \cos\varphi$ and $K = \sin\varphi$, $\varphi \in [0, 2\pi)$ [33]. The NN DMI is represented by the third, where $\vec{D}_{ij} = D\nu_{ij}\vec{e}_z$, $\vec{e}_z$ is the unit vector along $[111]$ direction and $\nu_{ij} = \pm 1$ correspond to the anticlockwise and clockwise directions of the NN couplings in a triangle, respectively. The last term is the energy of Zeeman coupling to the magnetic field $B$.

Generally, in the kagome magnet the Kitaev interaction does not support ferromagnetic order, it is necessary to add the final term of Zeeman coupling to an applied magnetic field along the [111] direction, which is strong enough to bring the system into fully polarized paramagnetic phase. In our work we set $h = 10$. Furthermore since there are complicated phases and phase transitions in the range $D \in [0, 0.02]$ and the magnitude of the related thermal conductivity is nearly zero, therefore, we will not study physical phenomena in this range in our study.

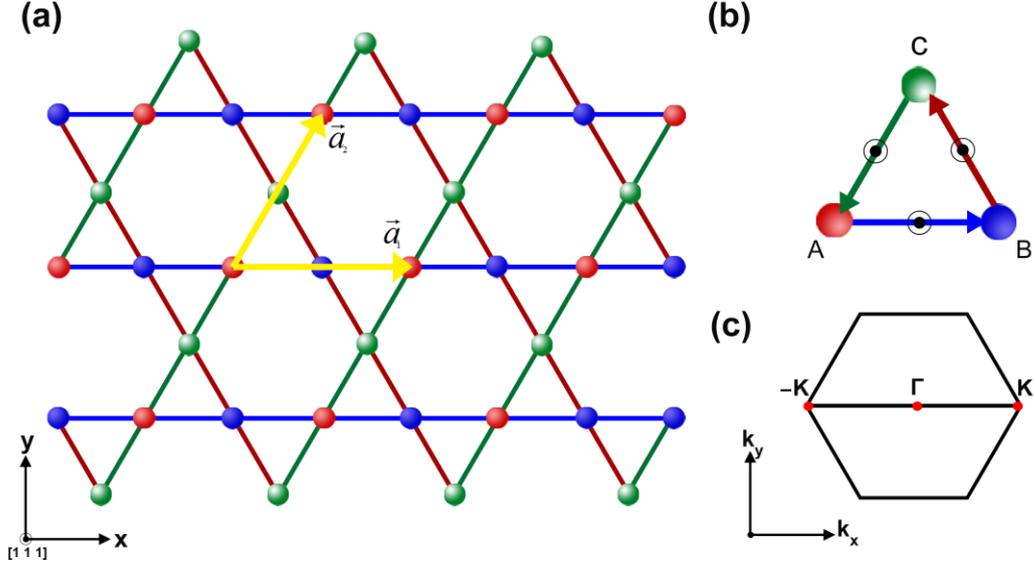

**Fig. 1** The schematics of the structure of kagome lattice. (a) The kagome lattice sits on the (111) plane and basis vectors are labeled as $a_1 = (1, 0)$ and $a_2 = (1, \sqrt{3})/2$. The red, green, and blue line represent bond links of Kitaev interaction. (b) The arrows in a triangle denote the coupling directions whose DM vectors are in the [111] direction. (c) The first Brillouin zone of the kagome lattice with two inequivalent high symmetry points at $K$ and $-K$.

We will restrict our study to linear spin wave theory, where we neglect interactions between magnons and only consider linear terms in the boson Hamiltonian, the linear term represents the fundamental and undisturbed behavior of the magnon. This approximation is reasonable and very efficient in prediction of the corresponding physics [34]. In the spin wave theory basis, we then employ the

Holstein-Primakoff transformation [35]: $S^z = S - a^\dagger a$, $S^+ = \sqrt{2S - a^\dagger a}\, a$, $S^- = a^\dagger \sqrt{2S - a^\dagger a}$; $a^\dagger$ is the boson creation operator. Upon inserting the Holstein-Primakoff transformation into the Hamiltonian, and after subsequent Fourier transformation, we obtain the linear spin-wave theory Hamiltonian, denoted as $H(\vec{k})$, It is a $2n \times 2n$ matrix, where $n = 3$.

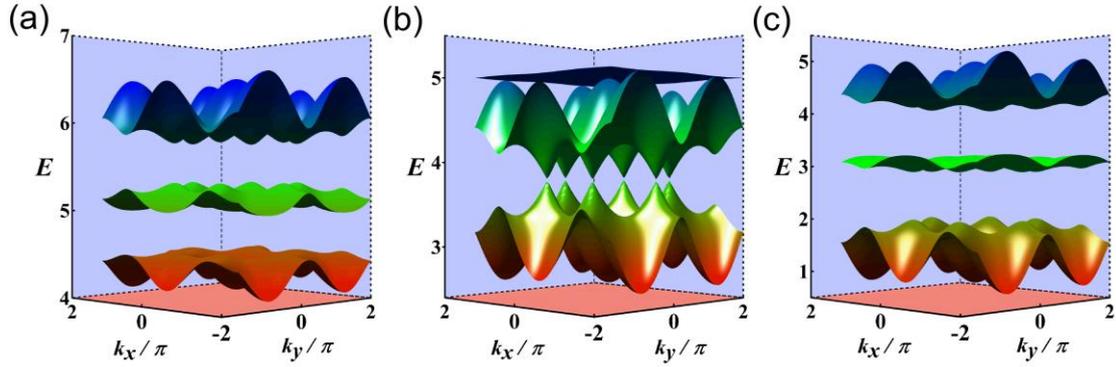

**Fig. 2**. Topological magnon band structure of kagome quantum ferromagnetic spin system. The corresponding parameters are as follows (a) $\varphi = 1.57\pi$, (b) $\varphi = 1.726\pi$, (c) $\varphi = 2\pi$ all with $D = 0.7$

We diagonalize the dynamical matrix of $H(\vec{k})$ based on the commutation relation $i\dfrac{d\Phi(\vec{k})}{dt} = [\Phi(\vec{k}), H(\vec{k})] = D'\Phi(\vec{k})$, where the dynamical matrix is given by $D' = \hat{g} H(\vec{k})$ with $\hat{g} = [(I, 0), (0, -I)]$, $I$ as the $3 \times 3$ identity matrix, and a basis is chosen as $\Phi(\vec{k}) = \left[a_{1,\vec{k}}^\dagger, a_{2,\vec{k}}^\dagger, a_{3,\vec{k}}^\dagger, a_{1,-\vec{k}}, a_{2,-\vec{k}}, a_{3,-\vec{k}}\right]^T$. Only positive real eigenvalues of the dynamical matrix $D'$ are considered and the stability of the system is confirmed when there are two non-negative eigenvalues for each vector $\vec{k}$. In Fig. 2 we show the Floquet magnon bands for undriven insulating kagome ferromagnets for $\varphi = 1.57\pi$ $\varphi = 1.726\pi$ and $\varphi = 2\pi$ all with $D = 0.7$. For $\varphi = 1.726\pi$. We can see that all the bands are separated by a finite energy gap when $\varphi = 1.726\pi$ and $\varphi = 2\pi$.

## 3. Periodically driven topological magnon insulators

The purpose of this paper is to periodically drive the magnon topologically trivial phases of eq. (1) to Floquet topological magnon insulators for $\varphi = 1.57\pi$ $\varphi = 1.726\pi$ and $\varphi = 2\pi$ with $D = 0.7$. We introduce the notion of periodically driven intrinsic topological magnon insulators. In this section, we introduce the notion of periodically driven intrinsic topological magnon insulators. Essentially, this concept will be based on the charge-neutrality of magnons in combination with their magnetic dipole moment $\vec{\mu} = \mu \hat{z}$ with $\vec{\mu} = g\mu_B$. Let us suppose that magnons in insulating quantum magnetic systems are exposed to an electromagnetic field with a dominant time-dependent electric field vector $\vec{E}(\tau)$. Then the effects of the field on the system can be described by a vector potential defined as $\vec{E}(\tau) = -\partial \vec{A}(\tau)/\partial \tau$, where $\vec{A}(\tau) = A[\sin(\omega\tau), \sin(\omega\tau + \phi), 0]$. The vector potential has time-periodicity: $\vec{A}(\tau + T) = \vec{A}(\tau)$, with $T = 2\pi/\omega$ being the period. Here $\phi = \pi/2$ corresponds to circularly-polarized light, whereas $\phi = 0$ corresponds to linearly-polarized light. In this study we only consider the case when $\phi = \pi/2$. We consider magnon quasiparticles with a magnetic dipole $\mu$ traveling in the backdrop of a time-dependent electric field, using the AC effect for charge-neutral particles. They will obtain a time-dependent AC phase in this case provided by $A_{ij}(\tau) = \mu \int_{\vec{r}_i}^{\vec{r}_j} \vec{A}(\tau) \cdot d\vec{l}$. The time-dependent Peierls substitution enables for the concise expression of the periodically driven magnon Hamiltonian $H(\tau)$.

We can now implement the machinery of the Floquet theory, to keep things simple, we'll focus on the magnonic Floquet Hamiltonian in the off-resonant regime, which occurs when the driving frequency ($\omega$) exceeds the magnon band-width of the undriven system ($\Delta$). In this case, the Floquet bands are decoupled, and it is sufficient to analyze the zeroth order time-independent Floquet magnon Hamiltonian $H_{eff}^0(\vec{k})$. The static effective Hamiltonian can be diagonalized by performing a bosonic Bogoliubov transformation.

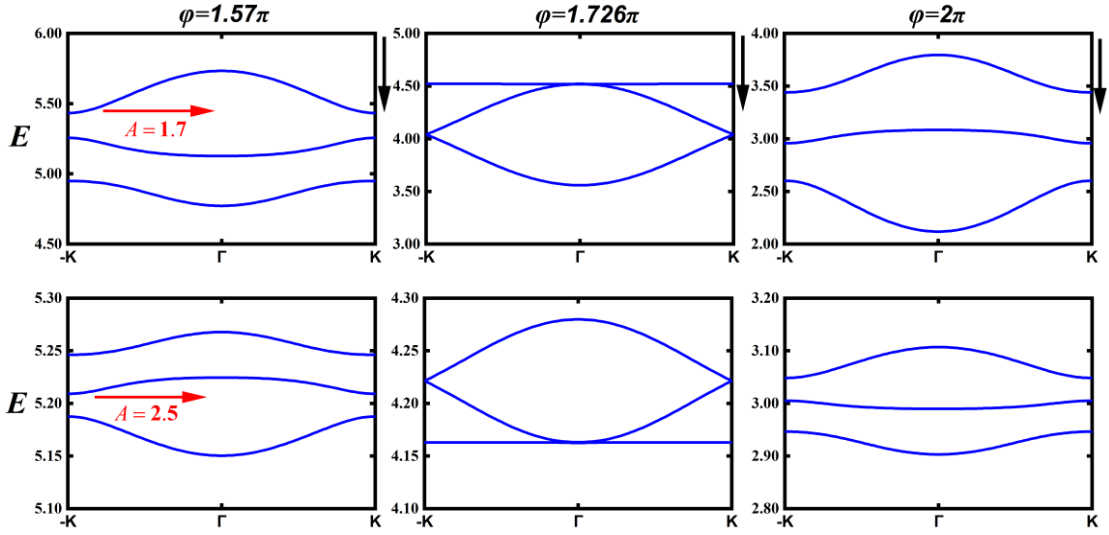

**Fig. 3** The Floquet magnon band in the Kitaev-Heisenberg kagome ferromagnet for $\varphi = 1.57\pi$, $\varphi = 1.726\pi$ and $\varphi = 2\pi$ all with $D = 0.7$

In Fig.3, we have shown magnon dispersions of Kitaev kagome magnet model for varying light intensity. We can see that the lower and upper magnetic oscillator bands and their corresponding Chern numbers change with light intensity, while the middle magnetic oscillator band remains constant. In the following, we shall discuss the topological aspects of this model. The Berry curvature is one of the main important quantities in topological systems. It is the basis of many observables in topological insulators. To study the photoinduced topological phase transitions in driven topological magnon insulators, we define the Berry curvature of a given magnon band

$$\Omega_\beta(\vec{k}) = -2\,\mathrm{Im} \sum_{\alpha \neq \beta} \frac{\left[\langle T_\beta(\vec{k})|\hat{v}_x|T_\alpha(\vec{k})\rangle \langle T_\alpha(\vec{k})|\hat{v}_y|T_\beta(\vec{k})\rangle\right]}{\left[\varepsilon_\beta(\vec{k}) - \varepsilon_\alpha(\vec{k})\right]^2} \tag{2}$$

where $\hat{v}_i = \partial H_{eff}^0(\vec{k})/\partial k_i\,(i=x,y)$ are the velocity operators, $T_\beta(\vec{k})$ are the magnon eigenvectors, and $\varepsilon_\beta(\vec{k})$ are the magnon energy bands. The associated Chern number [36-38] is defined as the integration of the Berry curvature over the first Brillouin zone

$$C_\beta = \frac{1}{2\pi} \int_{BZ} d^2k\, \Omega_\beta^z(\vec{k}) \tag{3}$$

where $\beta=1, 2, 3$ label the lower, middle, and upper magnon bands respectively.

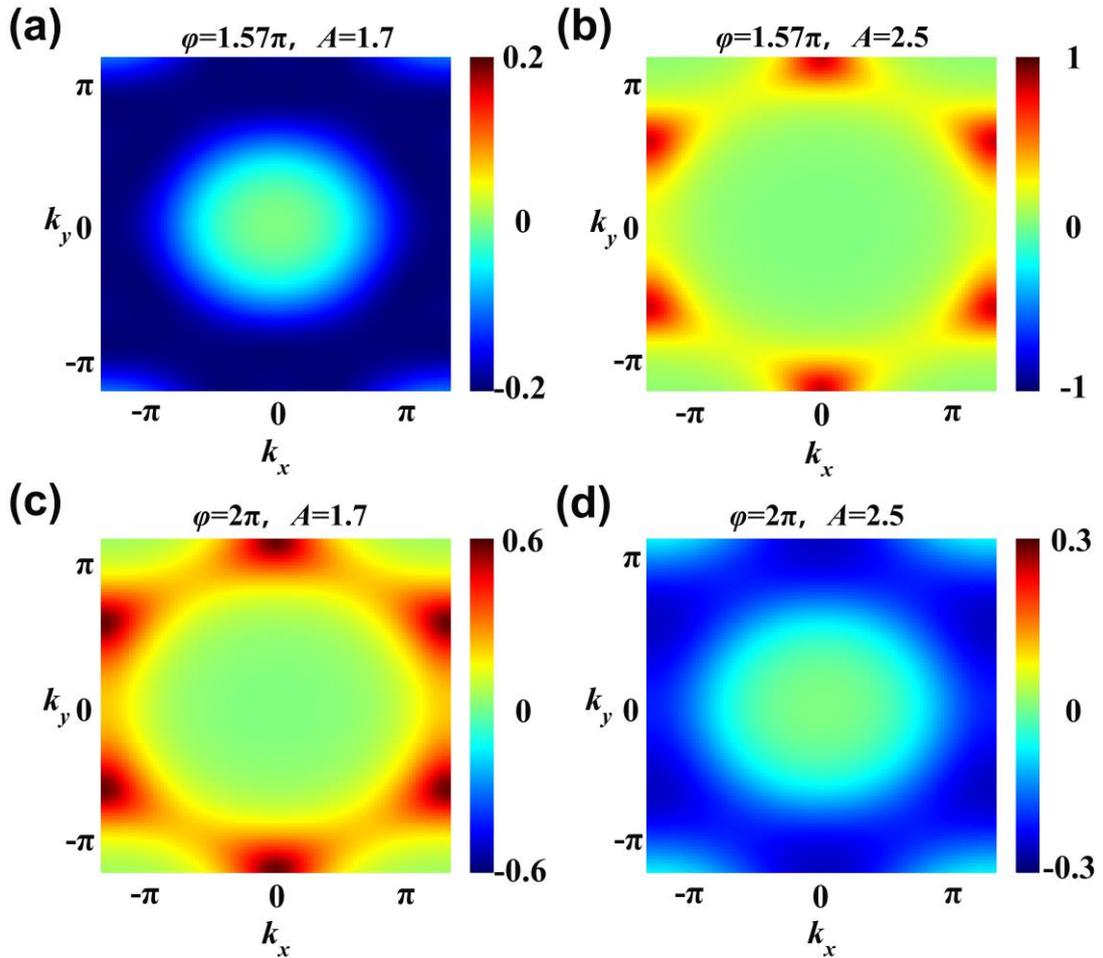

**Fig. 4** Berry curvature for the first magnon bands corresponding to Fig. 1 respectively. This leads to Chern number $C_\beta = \mp 1$.

In Fig. 4 we have shown the Berry curvatures for varying light intensity. We calculate the Berry curvatures of first magnon band of $\varphi = 1.726\pi$ and $\varphi = 2\pi$ with $D = 0.7$ respectively. When $A < A_0 \cong 2.41$, $\varphi = 1.726\pi$ and $\varphi = 2\pi$ have positive and negative Berry curvature corresponding to their first magnon bands, respectively. When $A > A_0 \cong 2.41$, the Berry curvature corresponding to the first magnon band of both has a sign reversal.

After calculating the Chern number as a fuction of $\varphi$ and $D$, we can know that here topological phase transition is found in the full region of the phase diagram as shown in Fig. 5.

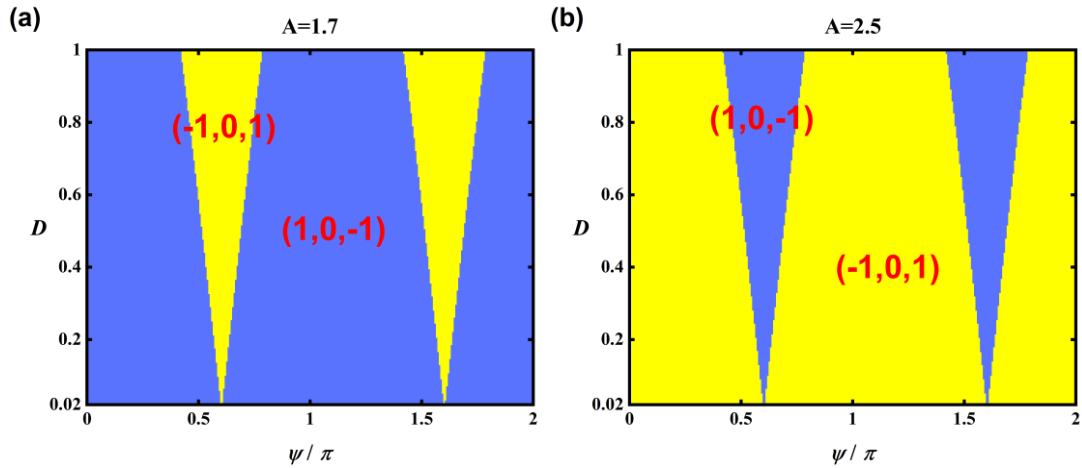

**Fig. 5** The topological phase diagram of the model as a function of $\varphi$ and $D$ for (a) $A = 1.7$, (b) $A = 2.5$.

To gain a deeper insight of the topological property, we calculate the Chern numble of the first magnon bands as a function of $A$ for $\varphi = 1.726\pi$ and $\varphi = 2\pi$, as shown in Fig. 6. The Chern number has been computed using the discretized Brillouin zone (BZ) method [39]. It is evident that by raising the light intensity, the current kagome ferromagnet switches from one Floquet topological magnon insulator with Chern numbers of $(-1,0,1)$ to another one with Chern numbers of $(1,0,-1)$. In other words, exposing a topological magnon insulator to a varying light intensity field will redistribute the magnon band structures, but both of them lead to a topological phase transition from one topological magnon insulator to another with different Berry curvatures and Chern numbers at a same light intensities $A_0=2.41$.

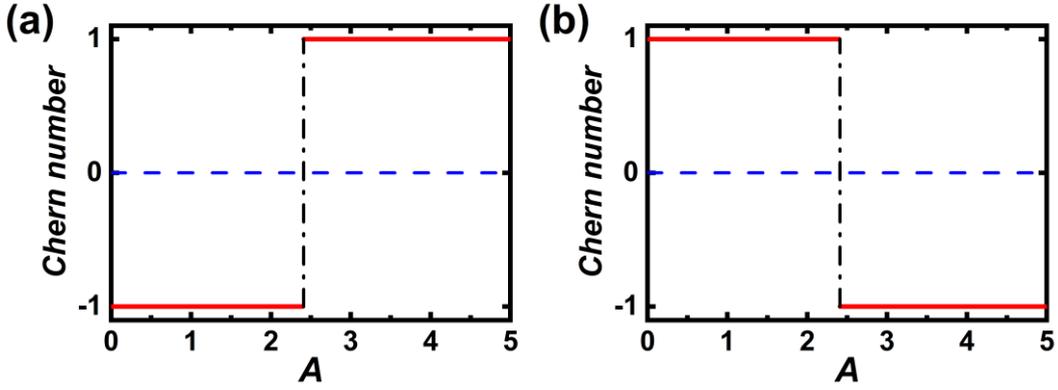

**Fig. 6** The Chern number of the first Floquet topological magnon band as a function of light intensities $A$ in the ferromagnetic Kitaev kagome ferromagnet model. (a) $\varphi=1.726\pi$, (b) $\varphi=2\pi$ both with $D=0.7$.

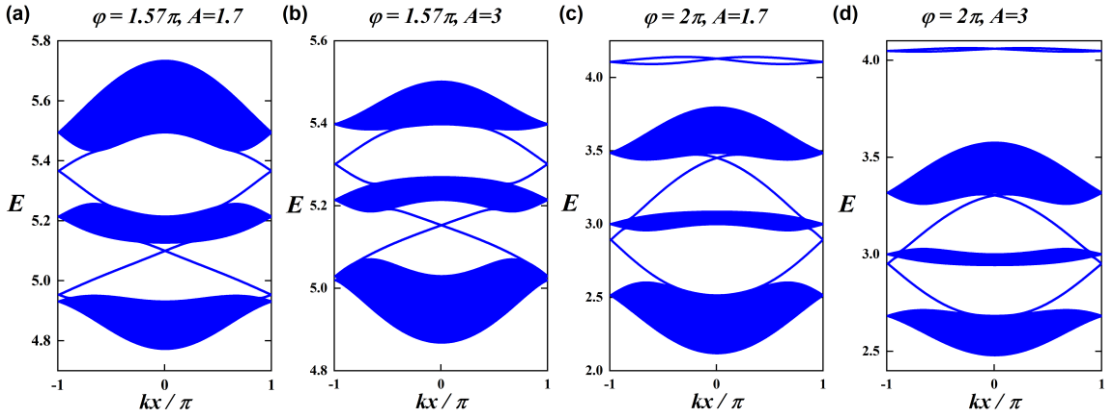

**Fig. 7** The corresponding magnon bands in a strip geometry under different parameters.

Due to the bulk-edge correspondence, the nonzero Chern numbers will promise in-gap edge modes in a strip geometry as shown in Fig. 7. Each diagram clearly shows that there are topologically protected crossed chiral edge states in each band gap. Furthermore, we note that there always exists Tamm-like edge states [40] in Fig. 7(a) and Fig. 7(b). Because they correspond to topologically trivial edge modes, which are not of interest in the present work, so we do not discuss it.

## 3. Thermal Hall effect

The presence of a finite Berry curvature in the magnon bands implies the existence of a thermal Hall effect. The thermal Hall effect in quantum magnets was first predicted theoretically by Katsura-Nagaosa-Lee [41]. Physically, we usually use the thermal Hall conductivity to characterize the strength of the thermal Hall effect, which in magnon systems is similar to Hall conductivity in electronic systems and is related to the Berry curvature of the eigenstates. The thermal Hall conductivity can be expressed as:

$$\kappa_{TH}^{xy} = -\frac{k_B^2 T}{(2\pi)^2 \hbar} \sum_{\beta=\pm} \int d^2k \, c_2(n_\beta) \Omega_\beta(\vec{k}), \quad (4)$$

where $n_\beta$ is the Bose-Einstein distribution function, $\Omega_\beta(\vec{k})$ is the Berry curvature of the $\beta$ th magnon band, $c_2(x) = (1+x)\left(\ln\frac{1+x}{x}\right)^2 - (\ln x)^2 - 2\text{Li}_2(-x)$ is a distribution function of the bosons, and $\text{Li}_2(x)$ represents the dilogarithm function.

In ferromagnetic insulators, researchers focus on the thermal Hall effect mainly on the topological magnon insulation phase, because in this special phase state, magnetons show extraordinary properties. Specifically, ferromagnetic insulators exhibit special thermal Hall effects when the lowest energy magnon band can be effectively separated from other energy bands, and these magnon bands carry a definite number of periods. For periodically driven magnon systems, we focus on there regime where the Bose distribution function is close to thermal equilibrium. In this regime, the same theoretical concept of the thermal Hall effect in undriven topological magnon systems can be applied to the driven magnon systems. Its dominant contribution comes from the peaks of the Berry curvature, Furthermore, the thermal Hall transport is mostly dictated by the lowest (acoustic) energy magnon band due to the bosonic structure of magnon.

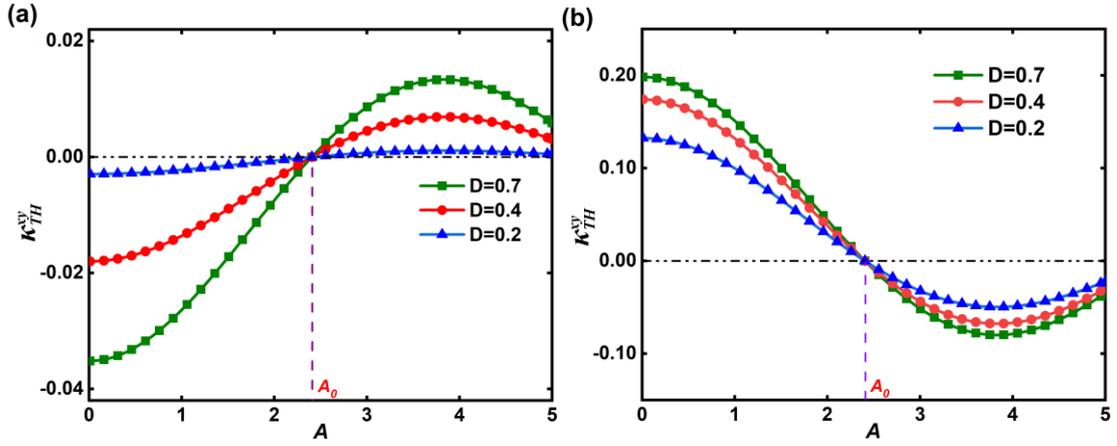

**Fig. 8** Tunable photoinduced Floquet thermal Hall conductivity $\kappa_{TH}^{xy}$ as a function of light intensities $A$ for (a) $\varphi = 1.726\pi$, (b) $\varphi = 2\pi$ both with $T = 1$.

In Fig. 8, we have shown the thermal Hall conductivity $\kappa_{TH}^{xy}$ as a function of the light intensitie $A$. The results show that the sign of the thermal Hall conductivity reverses after the light intensity is increased to a certain intensity. Moreover, the critical light intensity of the corresponding phase transition in different phases does not change with the value of $D$, and topological phase transitions occur at $A_0 = 2.41$, which is same as the light intensitie when the Chern number of the first magnon band changed. Because the thermal Hall conductivity of the phase transition boundary is close to zero, it is not discussed here. In Fig. 9, the thermal Hall conductivity as a function of temperature $T$ for different light intensities $A$ was illustrated, and the change of the symbol of the thermal Hall conductivity is used as a sign to judge the topological phase transition of the system. From the figure, we can clearly observe that with the change of temperature, the thermal Hall conductivity does not change symbolically, but the system has a topological phase transition with the increase of light intensity, and the critical light intensity for topological phase transition is $A_0 = 2.41$.

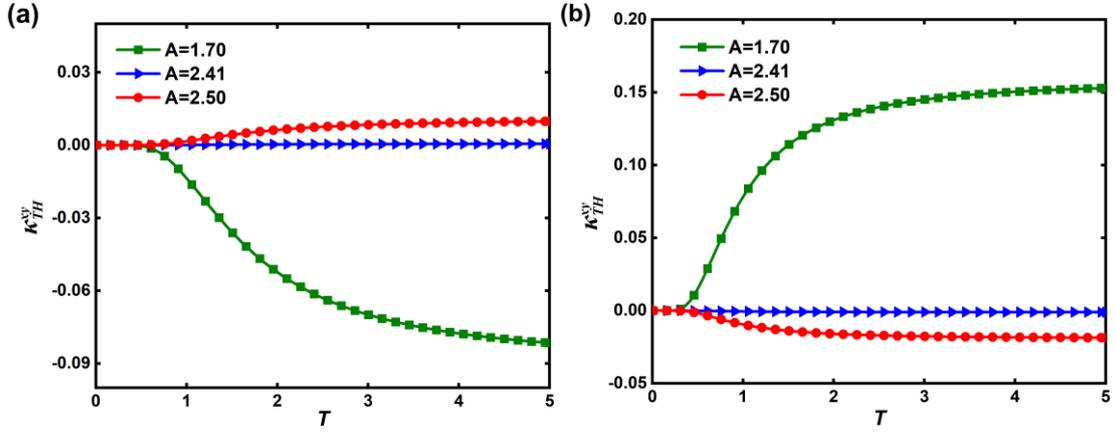

**Fig. 9** Tunable photoinduced Floquet thermal Hall conductivity $\kappa_{TH}^{xy}$ as a function of temperature $T$ for (a) $\varphi = 1.726\pi$, (b) $\varphi = 2\pi$ both with $D = 0.7$.

## 4. Summary

Our research focuses on photoinduced topological phase transitions in kagome magnet which is exhibiting Heisenberg, Kitaev and DM interactions and also exposed to a magnetic field by using the Floquet-Bloch theory to analyze. Using this theory, we show that , and thermal Hall conductivity. We have studied the off-resonant regime, when the driving frequency $\omega$ is larger than the magnon band-width $\Delta$ of the undriven system. In this regime, the Floquet sidebands are decoupled and can be considered independently. By observing the magnon dispersion curve of Kitaev kagome magnet model for varying light intensity, we know that the lower and upper magnon bands and their corresponding Chern numbers change with light intensity, while the middle magnetic oscillator band remains constant. The main result of this report is that intrinsic topological magnon insulators in the kagome magnets can be driven from one topological magnon insulators to another with different Berry curvature and Chern number by manipulating the light intensity.

Even exposing a topological magnon insulator to a varying light intensity field which will redistribute the magnon band structures, both of them will lead to a topological phase transition from one topological magnon insulator to another with different Berry curvatures and Chern numbers at a same light intensities. Also, the sign of the thermal Hall conductivity reverses after the light intensity is increased to a certain intensity $A_0 = 2.41$, since each topological phase is connected with a distinct sign of the thermal Hall conductivity $\kappa_{TH}^{xy}$. Although the value of $D$ will influence the magnitude of the change in the the thermal Hall conductivity, the critical light intensity of the corresponding phase transition in different phases does not change with the value of $D$, which is all same as the light intensitie when the Chern number of the first magnon band changed. Finally, we hope that our theoretical ideas can be realized in realistic materials.

**Acknowledgments**

This work was supported by the National Natural Science Foundation of China under Grant No.12064011, 12375014, 11875149, the Natural Science Fund Project of Hunan Province under Grant No. 2020JJ4498 and the Graduate Research Innovation Foundation of Jishou University under Grant No. Jdy21030.